\documentstyle[editedvolume]{crckapb}

\begin{opening}
\title{The Wave Mechanics of Large-scale Structure}

\author{Peter Coles}
\institute{School of Physics \& Astronomy,\\
           University of Nottingham, \\
           University Park,\\
           Nottingham NG7 2RD,\\
           United Kingdom}

\end{opening}

\runningtitle{Wave Mechanics of Large-scale Structure}

\begin{document}

\begin{abstract}
I review the basic ``gravitational instability'' model for the
growth of structure in the expanding Universe. This model requires
the existence of small initial irregularities in the density of a
largely uniform universe. These grow through linear and non-linear
stages to form a complex network of clusters, filaments and voids.
The dynamical equations describing the evolution of a
self-gravitating fluid can be rewritten in the form of a
Schr\"{o}dinger equation coupled to a Poisson equation determining
the gravitational potential. This approach has a number of
interesting features, many of which were pointed out in a seminal
paper by Widrow \& Kaiser (1993). I argue that this approach has
the potential to yield useful analytic insights into the dynamical
growth of large-scale structure. As a particular example, I show
that this approach yields an elegant reformulation of an idea due
to Jones (1999) concerning the origin of lognormal intermittency
in the galaxy distribution.
\end{abstract}

\section{Introduction}

The local Universe displays a rich hierarchical pattern of galaxy
clustering that encompasses a large range of length scales,
culminating in rich clusters and superclusters. The early
Universe, however, was almost smooth, with only slight ripples
seen in the cosmic microwave background radiation. Models of the
evolution of structure link these observations through the effect
of gravity, because the small initially overdense fluctuations
accrete additional matter as the Universe expands. During the
early stages, the ripples evolve independently, like linear waves
on the surface of deep water, but as the structures grow in mass,
they interact with other in non-linear ways, more like nonlinear
waves breaking in shallow water.

The linear theory of perturbation growth is well-established, but
the non-linear regime is much more complicated and generally not
amenable to analytic solution. Numerical $N$--body simulations
have led the way towards an understanding of strongly developed
clustering, but simulating a thing is not quite equivalent to
understanding it. In this lecture, therefore, I sketch out some of
the analytic methods that can be used to study non-linear
clustering. I focus in particular on a novel approach based on the
description of density fluctuations using quantum mechanics; see
also Coles (2002).

\section{Cosmological Structure Formation}

The Big Bang theory is built upon the Cosmological Principle, a
symmetry principle that requires the Universe on large scales to
be both homogeneous and isotropic. Space-times consistent with
this requirement can be described by the Robertson--Walker metric
\begin{equation}
{\rm d}s_{\rm FRW}^2 = c^2 {\rm d}t^2 - a^2(t)\left({{\rm
d}r^2\over 1 - \kappa r^2} + r^2 {\rm d}\theta^2 + r^2\sin^2\theta
{\rm d}\phi^2\right)  , \label{eq:l1a}
\end{equation}
where $\kappa$ is the spatial curvature, scaled so as to take the
values $0$ or $\pm 1$. The case $\kappa=0$ represents  flat space
sections, and the other two cases are  space sections of constant
positive or negative curvature, respectively. The time coordinate
$t$ is called {\em cosmological proper time} and it is singled out
as a preferred time coordinate by the property of spatial
homogeneity. The quantity $a(t)$,  the {\em cosmic scale factor},
describes the overall expansion of the universe  as a function of
time. If light emitted at time $t_{\rm e}$ is received by an
observer at $t_0$ then the redshift $z$ of the source is given by
\begin{equation}
1+z = \frac{a(t_0)}{a(t_{\rm e})}.
\end{equation}
The dynamics of an FRW universe are determined by the Einstein
gravitational field equations which become
\begin{eqnarray}
3\left( \frac{\dot{a}}{a} \right)^{2} & = & 8\pi G\rho - {3\kappa
c^{2} \over a^2} + \Lambda,\\ {\ddot{a}\over a} & = & - {4\pi
G\over 3} \left(\rho + 3 \frac{p}{c^2}\right) + {\Lambda\over 3},
\\ \dot{\rho}& =& - 3 {\dot{a}\over a}\left(\rho + \frac{p}{c^2}
\right). \label{eq:l1b}
\end{eqnarray}
These equations  determine the time evolution of the cosmic scale
factor $a(t)$ (the dots denote derivatives with respect to
cosmological proper time $t$) and therefore describe the global
expansion or contraction of the universe. The behaviour of these
models can further be parametrised in terms of the Hubble
parameter $H=(\dot{a}/a)$ and the density parameter $\Omega=8\pi
G\rho/3H^2$, a suffix $0$ representing the value of these
quantities at the present epoch when $t=t_0$.

In order to understand how of large-scale structure arises, it is
best to begin with the standard fluid-based approach to structure
growth. In the standard treatment of the Jeans Instability one
begins with the dynamical equations governing the behaviour of a
self-gravitating fluid. These are the {\em Euler equation}
\begin{equation}
{\partial ({\bf v})\over \partial t} + ({\bf v}\cdot{\bf
\nabla}){\bf v}  + {1\over \rho}{\bf \nabla} p + {\bf
\nabla}\phi=0~; \label{eq:Euler1}
\end{equation}
the {\em continuity equation} \begin{equation} {\partial\rho\over
\partial t} +  +  {\bf \nabla} (\rho{\bf v}) =
0~, \label{eq:continuity1}
\end{equation}
expressing the conservation of matter; and the {\em Poisson
equation}
\begin{equation} {\bf
\nabla}^2\phi = 4\pi G \rho~, \label{eq:Poisson1}
\end{equation}
describing Newtonian gravity. If the length scale of the
perturbations is smaller than the effective cosmological horizon
$d_H=c/H$, a Newtonian treatment of cosmic structure formation is
still expected to be valid in expanding world models. In an
expanding cosmological background, the Newtonian equations
governing the motion of gravitating particles can be written in
terms of
\begin{equation}
 {\bf x} \equiv {\bf r} / a(t)\end{equation} (the comoving spatial
coordinate, which is fixed for observers moving with the Hubble
expansion),
\begin{equation} {\bf v} \equiv \dot {{\bf r}} - H {\bf r} = a\dot
{{\bf x}}\end{equation} (the peculiar velocity field, representing
departures of the matter motion from pure Hubble expansion), $\rho
({\bf x}, t)$ (the matter density), and $\phi ({\bf x} , t)$ (the
peculiar Newtonian gravitational potential, i.e. the fluctuations
in potential with respect to the homogeneous background)
determined by the Poisson equation in the form
\begin{equation}
{\bf \nabla_x}^2\phi = 4\pi G a^2(\rho - \rho_0) = 4\pi
Ga^2\rho_0\delta. \label{eq:Poisson}
\end{equation}
In this equation and the following the suffix on $\nabla_x$
indicates derivatives with respect to the new comoving
coordinates. Here $\rho_0$ is the mean background density, and
\begin{equation}
\delta \equiv \frac{\rho-\rho_0}{\rho_0}
\end{equation}
is the {\em density contrast}. Using these variables the Euler
equation becomes
\begin{equation}
{\partial (a{\bf v})\over \partial t} + ({\bf v}\cdot{\bf
\nabla_x}){\bf v} = - {1\over \rho}{\bf \nabla_x} p - {\bf
\nabla_x}\phi~. \label{eq:Euler}
\end{equation}
The first term on the right-hand-side of equation (\ref{eq:Euler})
arises from pressure gradients, and is neglected in dust-dominated
cosmologies. Pressure effects may nevertheless be important in the
the (collisional) baryonic component of the mass distribution when
nonlinear structures eventually form. The second term on the
right-hand side of equation (\ref{eq:Euler}) is the peculiar
gravitational force, which can be written in terms of ${\bf g} =
-{\bf \nabla_x}\phi/a$, the peculiar gravitational acceleration of
the fluid element. If the velocity flow is irrotational, ${\bf v}$
can be rewritten in terms of a velocity potential $\phi_v$:
\begin{equation}{\bf v} = - {\bf
\nabla_x} \phi_v/a. \end{equation} This is expected to be the case
in the cosmological setting because (a) there are no sources of
vorticity in these equations and (b) vortical perturbation modes
decay with the expansion. We also have a revised form of the
continuity equation:
\begin{equation}
{\partial\rho\over \partial t} + 3H\rho + {1\over a} {\bf
\nabla_x} (\rho{\bf v}) = 0~. \label{eq:continuity}
\end{equation}

In order to understand how structures form we need to consider the
difficult problem of dealing with the evolution of inhomogeneities
in the expanding Universe. We are helped in this task by the fact
that we expect such inhomogeneities to be of very small amplitude
early on so we can adopt a kind of perturbative approach, at least
for the early stages of the problem. The procedure is  to
linearise the Euler, continuity and Poisson equations by
perturbing physical quantities defined as functions of Eulerian
coordinates, i.e. relative to an unperturbed coordinate system.
Expanding $\rho$, ${\bf v}$ and $\phi$ perturbatively and keeping
only the first-order terms in equations (\ref{eq:Euler}) and
(\ref{eq:continuity}) gives the linearised continuity equation:
\begin{equation}
{\partial\delta\over \partial t} = - {1\over a}{\bf \nabla_x}\cdot
{\bf v},
\end{equation}
which can be inverted, with a suitable choice of boundary
conditions, to yield
\begin{equation}
\delta = - {1\over a H f}\left({\bf \nabla_x}\cdot{\bf v}\right).
\label{eq:l9}
\end{equation}
The function $f\simeq \Omega_0^{0.6}$; this is simply a fitting
formula to the full solution (Peebles 1980). The linearised Euler
and Poisson equations are
\begin{equation}
{\partial {\bf v}\over\partial t} + {\dot a\over a}{\bf v} = -
{1\over \rho a}{\bf \nabla_x} p -{1\over a}{\bf \nabla_x}\phi,
\label{eq:l10}
\end{equation}
\begin{equation}
{\bf \nabla_x}^2\phi = 4\pi G a^2\rho_0\delta; \label{eq:l11}
\end{equation}
$|{\bf v}|, |\phi|, |\delta| \ll 1$ in equations (\ref{eq:l9}),
(\ref{eq:l10}) \& (\ref{eq:l11}). From these equations, and if one
ignores pressure forces, it is easy to obtain an equation for the
evolution of $\delta$:
\begin{equation}
\ddot\delta + 2H\dot\delta - {3\over 2}\Omega H^2\delta = 0.
\label{eq:l13b}
\end{equation}
For a spatially flat universe dominated by pressureless matter,
$\rho_0(t) = 1/6\pi Gt^2$ and equation (\ref{eq:l13b}) admits two
linearly independent power law solutions $\delta({\bf x},t) =
D_{\pm}(t)\delta({\bf x})$, where $D_+(t) \propto a(t) \propto
t^{2/3}$  is the growing  mode and $D_-(t) \propto t^{-1}$ is the
decaying mode.

The above considerations apply to the evolution of a single
Fourier mode of the density field $\delta({\bf x}, t) =
D_+(t)\delta({\bf x})$. What is more likely to be relevant,
however, is the case of a superposition of waves, resulting from
some kind of stochastic process in which he density field consists
of a  superposition of such modes with different amplitudes. A
statistical description of the initial perturbations is therefore
required, and any comparison between theory and observations will
also have to be statistical. Many versions of the inflationary
scenario for the very early universe (Guth 1981; Guth \& Pi 1982;
Brandenberger 1985) predict the initial density fluctuatiosn to
take the form of a {\em Gaussian random field} in which the
initial Fourier modes of the perturbation field have random
phases.

\section{Nonlinear Gravitational Instability}

The linearised equations of motion  provide an excellent
description of gravitational instability at very early times when
density fluctuations are still small ($\delta \ll 1$). The linear
regime of gravitational instability breaks down when $\delta$
becomes comparable to unity, marking the commencement of the {\it
quasi-linear} (or weakly non-linear) regime. During this regime
the density contrast may remain small ($\delta < 1$), but the
phases of the Fourier components $\delta_{\bf k}$ become
substantially different from their initial values resulting in the
gradual development of a non-Gaussian distribution function if the
primordial density field was Gaussian. In this regime the shape of
the power-spectrum changes by virtue of a complicated cross-talk
between different wave-modes. Analytic methods are available for
this kind of problem (Sahni \& Coles 1995), but the usual approach
is to use $N$-body experiments for strongly non-linear analyses
(Davis et al. 1985; Jenkins et al. 1998).

Further into the non-linear regime, bound structures form. The
baryonic content of these objects may then become important
dynamically: hydrodynamical effects (e.g. shocks), star formation
and heating and cooling of gas all come into play. The spatial
distribution of galaxies may therefore be very different from the
distribution of the (dark) matter, even on large scales. Attempts
are only just being made to model some of these processes with
cosmological hydrodynamics codes, such as those based on Smoothed
Particle Hydrodynamics (SPH; Monaghan 1992), but it is some
measure of the difficulty of understanding the formation of
galaxies and clusters that most studies have only just begun to
attempt to include modelling the detailed physics of galaxy
formation. In the front rank of theoretical efforts in this area
are the so-called semi-analytical models which encode simple rules
for the formation of stars within a framework of merger trees that
allows the hierarchical nature of gravitational instability to be
explicitly taken into account.

Perturbation theory fails when nonlinearities develop but it is
important to stress that the fluid treatment is intrinsically
approximate anyway. A fuller treatment of the problem requires a
solution of the Boltzmann equation for the full phase-space
distribution of the system $f({\bf x}, {\bf v}, t)$ coupled to the
Poisson equation (\ref{eq:Poisson1}) that determines the
gravitational potential. In cases where the matter component is
collisionless, the Boltzmann equation takes the form of a Vlasov
equation:
\begin{equation}
{\partial f \over \partial t}= \sum_{i=1}^{3} \left({\partial
\phi\over
\partial x_i}{\partial f \over \partial v_i} - v_i {\partial f \over \partial
x_i}\right).
\end{equation}
The fluid approach  outline above can only describe cold material
where the velocity dispersion of particles is negligible. But even
if the dark matter is cold, there may be hot components of
baryonic material whose behaviour needs also to be understood.
Moreover, the fluid approach assumes the existence of a single
fluid velocity at every spatial position. It therefore fails when
orbits cross and multi-streaming generates a range of particle
velocities through a given point.

Fortunately the formation of these structural elements can also be
understood using simple models, especially that of Zel'dovich
(1970). This approximation actually predicts that the density in
certain regions -- called {\it caustics} -- should become
infinite, but the gravitational acceleration caused by these
regions remains finite. Of course, in any case one cannot justify
ignoring pressure when the density becomes very high, for much the
same reason as we discussed above  in the context of spherical
collapse: one forms shock waves which compress infalling material.
At a certain point the process of accretion onto the caustic will
stop: the condensed matter is contained by gravity within the
final structure, while the matter which has not passed through the
shock wave is held up by pressure. It has been calculated that
about half the material inside the original fluctuation is
reheated and compressed by the shock wave. An important property
of the structures which thus form is that they are strongly
unstable to fragmentation. In principle, therefore, one can
generate structure on smaller scales than the pancake.

I will now describe the Zel'dovich approximation in more detail,
and show how it can follow the evolution of perturbations until
the formation of pancakes. Imagine that we begin with a set of
particles which are uniformly distributed in space. Let the
initial (i.e. Lagrangian) coordinate of a particle in this
unperturbed distribution be ${\bf q}$. Now each particle is
subjected to a displacement corresponding to a density
perturbation. In the Zel'dovich approximation the  Eulerian
coordinate of the particle at time $t$ is  \begin{equation}{\bf r}
(t, {\bf q}) = a(t) [{\bf q}-b(t) {\bf \nabla}_{\bf q} \Phi_0
({\bf q})], \end{equation} where ${\bf r}=a(t){\bf x}$, with ${\bf
x}$ a comoving coordinate, and we have made $a(t)$ dimensionless
by dividing throughout by $a(t_i)$, where $t_i$ is some reference
time which we take to be the initial time. The derivative on the
right hand side is taken with respect to the Lagrangian
coordinates. The dimensionless function $b(t)$ describes the
evolution of a perturbation in the linear regime, with the
 condition $b(t_i)=0$, and therefore solves the equation
\begin{equation} \ddot {b} +2 \left({\dot a \over a}\right) \dot b- 4 \pi G \rho b =
0~;
\end{equation}
cf. equation (20). For a flat matter--dominated universe we have
 $b \propto t^{2/3}$ as before. The quantity
$\Phi_0({\bf  q})$ is proportional  to a velocity potential, of
the type introduced above, i.e. a quantity of which the velocity
field is the gradient:
\begin{equation}
{\bf V}= {d{\bf r} \over dt} - H {\bf r} = a {d {\bf x} \over dt}
= - a \dot b {\bf \nabla}_q \Phi_0({\bf q});
\end{equation} this means that the velocity field is irrotational. The
quantity $\Phi_0 ({\bf q})$ is related to the density perturbation
in the linear regime by the relation $$ \delta = b \nabla^2 _{\bf
q} \Phi_0,$$ which is a simple consequence of Poisson's equation.

The Zel'dovich approximation is a linear approximation with
respect to the particle displacements rather than the density, as
was the linear solution we derived above. It is conventional to
describe the Zel'dovich approximation as a kind of first-order
Lagrangian perturbation theory, while what we have dealt with so
far for $\delta(t)$ is a first order Eulerian theory. We have also
assumed that the position and time dependence of the displacement
between initial and final positions can be separated. Notice that
particles in the Zel'dovich approximation execute a kind of
inertial motion on straight line trajectories.

The Zel'dovich approximation, though simple, has a number of
interesting properties. First, it is exact for the case of one
dimensional perturbations up to the moment of shell crossing. As
we have mentioned above, it also incorporates irrotational motion,
which is required to be the case if it is generated only by the
action  of gravity (due to the Kelvin circulation theorem). For
small displacements between ${\bf r}$ and $a(t){\bf q}$, one
recovers the usual (Eulerian) linear regime: in fact, equation
(22) defines a unique mapping between the coordinates ${\bf q}$
and ${\bf r}$ (as long as trajectories do not cross); this means
that
 $\rho ({\bf r} ,t) d^3 r =
\langle \rho(t_i) \rangle d^3 q$ or \begin{equation} \rho({\bf r},
t ) = { \langle \rho (t)\rangle \over \vert J ({\bf r}, t) \vert
}~,\end{equation} where $\vert J  ({\bf r}, t) \vert$ is the
determinant of the Jacobian of the mapping between ${\bf q}$ and
${\bf r}$: $\partial {\bf r} / \partial {\bf q}$.  Since the flow
is irrotational the matrix $J$ is symmetric and can therefore be
locally diagonalised. Hence \begin{equation}
 \rho({\bf r}, t ) =
\langle \rho (t)\rangle \prod_{i=1} ^3  [1 + b(t) \alpha_i({\bf
q})]^{-1}: \end{equation} the quantities $1 + b (t) \alpha_i$ are
the eigenvalues of the matrix $J$ (the $\alpha_i$ are the
eigenvalues of the deformation tensor). For times close to $t_i$,
when $\vert b(t) ~\alpha_i\vert \ll 1$, equation (26) yields
\begin{equation}
 \delta \simeq - (\alpha_1 + \alpha_2 + \alpha_3) b(t),
\end{equation}
which is just the law of perturbation growth in the linear regime
written a different way.

At some time $t_{sc}$, when $ b(t_{sc}) = - 1/\alpha_j$, an event
called {\it shell--crossing} occurs such that  a singularity
appears and the density becomes formally infinite in a region
where at least one of the eigenvalues (in this case $\alpha_j$) is
negative. This condition corresponds to the situation where two
points with different Lagrangian coordinates end up at the same
Eulerian coordinate. In other words, particle trajectories have
crossed and the mapping between Lagrangian and Eulerian space is
no longer unique. A region where the shell--crossing occurs is
called a caustic. For a fluid element to be collapsing, at least
one of the $\alpha_j$ must be negative. If more than one is
negative, then collapse will occur first along the axis
corresponding to the most negative eigenvalue. If there is no
special symmetry, one therefore expects collapse to be generically
one--dimensional, from three dimensions to two. Only if two (or
three) negative eigenvalues, very improbably,  are equal in
magnitude can the collapse occur to a filament (or point). One
therefore expects the formation of flattened structures to be the
generic result of such  collapse. This is in accord with the
classic work of Lin, Mestel \& Shu (1965) who showed that, for a
generic triaxial perturbation, the collapse is expected to occur
not to a point, but to a flattened structure of quasi--two
dimensional nature. The usual descriptive term for such features
is {\it pancakes}.

The Zel'dovich approximation matches very well the evolution of
density perturbations in full $N$--body calculations until the
point where shell crossing occurs (Coles, Melott \& Shandarin
1993). After this, the approximation breaks down completely.
Particles continue to move through the caustic in the same
direction as they did before. Particles entering a pancake from
either side merely sail through it and pass out the opposite side.
The pancake therefore appears only instantaneously and is rapidly
smeared out. In reality, the matter in the caustic would feel the
strong gravity there and be pulled back towards it before it could
escape through the other side. Since the Zel'dovich approximation
is only kinematic it does not account for these close--range
forces and the behaviour in the strongly non--linear regime is
therefore described very poorly. Furthermore, this approximation
cannot describe the formation of shocks and phenomena associated
with pressure.

Attempts to understand properties of non-linear structure using
the fluid model therefore resort to further approximations (Sahni
\& Coles 1995) to extend the approach beyond shell-crossing. One
relatively straightforward way to extend the Zel'dovich
approximation is through the so--called {\it adhesion model}
(Gurbatov, Saichev \& Shandarin 1989). In this  model one assumes
that the particles stick to each other when they enter a caustic
region because of an artificial viscosity which is intended to
simulate the action of strong gravitational effects inside the
overdensity forming there. This `sticking' results in a
cancellation of the component of the velocity of the particle
perpendicular to the caustic. If the caustic is two--dimensional,
the particles will move in its plane until they reach a
one--dimensional interface between two such planes. This would
then form a filament. Motion perpendicular to the filament would
be cancelled, and the particles will flow along it until a point
where two or more filaments intersect, thus forming a node. The
smaller is the viscosity term, the thinner will be the sheets and
filaments, and the more point--like will be the nodes. Outside
these structures, the Zel'dovich approximation is still valid to
high accuracy. Comparing simulations made within this
approximation with full $N$--body calculations shows that it is
quite accurate for overdensities up to
 $ \delta  \simeq 10 $.

The spatial distribution of particles obtained using the adhesion
approximation represents a sort of ``skeleton'' of the real
structure: non--linear evolution generically leads to the
formation of a quasi--cellular structure, which is a kind of
``tessellation'' of irregular polyhedra having pancakes for faces,
filaments for edges and nodes at the vertices . This skeleton,
however, evolves continuously as structures merge and disrupt each
other through tidal forces; gradually, as evolution proceeds, the
characteristic scale of the structures increases. In order to
interpret the observations we have already described, one can
think of the giant ``voids'' as being the regions internal to the
cells, while the cell nodes correspond to giant clusters of
galaxies. While analytical methods, such as the adhesion model,
are useful for mapping out the skeleton of structure formed during
the non--linear phase, they are not adequate for describing the
highly non--linear evolution within the densest clusters and
superclusters. In particular, the adhesion model cannot be used to
treat the process of merging and fragmentation of pancakes and
filaments due to their own (local) gravitational instabilities,
which must be done using full numerical computations.

\section{An Alternative Approach}

A novel approach to the study of collisionless matter, with
applications to structure formation, was suggested by Widrow \&
Kaiser (1993). It iinvolves re-writing of the fluid equations
given in Section 2 in the form of a non-linear Schr\"{o}dinger
equation. The equivalence between this and the fluid approach has
been known for some time; see Spiegel (1980) for historical
comments. Originally the interest was to find a fluid
interpretation of quantum mechanical effects, but in this context
we shall use it to describe an entirely classical system.

To begin with, consider the continuity equation and Euler equation
for a curl-free flow in which ${\bf v}=\nabla \phi)$, in response
to some general potential $V$. In this case the continuity
equation can be written
\begin{equation}
\frac{\partial \rho}{\partial t} + \nabla \cdot (\rho\nabla\phi)=
0~. \end{equation} It is convenient to take the first integral of
the Euler equation, giving
\begin{equation}
\frac{\partial \phi}{\partial t} + \frac{1}{2} (\nabla \phi)^2 =
-V~,
\end{equation}
which is usually known as the Bernoulli equation. The trick then
is to make a transformation of the form
\begin{equation}
\psi=\alpha \exp(i\phi/\nu)~,
\end{equation}
where $\rho=\psi\psi^*=|\psi|^2=\alpha^2$; the wavefunction
$\psi({\bf x}, t)$ evidently complex. Notice that the dimensions
of $\nu$ are the same as $\phi$, namely $[L^2T^{-1}]$. After some
algebra it emerges that the two equations above can be written in
one equation of the form
\begin{equation}
i\nu \frac{\partial \psi}{\partial t} = -\frac{\nu^2}{2} \nabla^2
\psi + V\psi + P\psi~, \label{eq:scrap}
\end{equation}
where \begin{equation} P=\frac{\nu^2}{2} \frac{\nabla^2
\alpha}{\alpha}. \end{equation} To accommodate gravity we need to
couple  equation (\ref{eq:scrap}) to the Poisson equation in the
form
\begin{equation}
\nabla^2\phi=4\pi G\psi \psi^*~,
\end{equation}
taking $V$ to be $\phi$.

This system looks very similar to a Schr\"{o}dinger equation,
except for the extra ``operator'' $P$. The role of this term
becomes clearer if one leaves it out of equation (\ref{eq:scrap})
and works backwards. The result is an extra term in the Bernouilli
equation that resembles a pressure gradient. This is often called
the ``quantum pressure'' that arises when one tries to understand
a quantum system in terms of a classical fluid behaviour. Leaving
it out to model a collisionless fluid can be justified only if
$\alpha$ varies only slowly on the scales of interest. On the
other hand one can model situations in which one wishes to model
genuine effects of pressure by adjusting (or omitting) this term
in the wave equation. Widrow \& Kaiser (1993) advocated simply
writing
\begin{equation}
i\hbar {\partial \psi\over \partial t}= - {\hbar^2 \over 2m}
\nabla^2\psi + m \phi({\bf x})\psi,\label{eq:schrod}
\end{equation}
i.e. simply ignoring the quantum pressure term. In this spirit,
the constant $\hbar$ is taken to be an adjustable parameter that
controls the spatial resolution $\lambda$ through a de Broglie
relation $\lambda=\hbar/mv$. In terms of the parameter $\nu$ used
above, we have $\nu=\hbar/m$, giving the correct dimensions for
Planck's constant. Note that the wavefunction $\psi$ encodes the
velocity part of phase space in its argument through the ansatz
\begin{equation}
\psi({\bf x})=\sqrt{\rho({\bf x})} \exp [i\theta({\bf x}/\hbar)],
\end{equation}
where $\nabla \theta({\bf x})={\bf p}({\bf x})$, the local
`momentum field'. This formalism thus yields an elegant
description of both the density and velocity fields in a single
function.

The approach outlined above is relatively new to galaxy clustering
studies, and many details still need to be worked out. One source
of complexity arises when one places the system in an expanding
context. To see what happens, let us define a scaled density
$\chi=\rho/\rho_0=(1+\delta)$ and take $\Omega=1$. The continuity
equation then becomes
\begin{equation}
\frac{\partial \chi}{\partial a} + \nabla \cdot (\chi\nabla\phi)=
0~,
\end{equation}
where the velocity potential $\phi$ is now such that ${\bf
u}=d{\bf x}/dt=\nabla\phi$ and $a$ is the scale factor. It is
convenient to take the first integral of the Euler equation,
giving
\begin{equation}
\frac{\partial \phi}{\partial a} + \frac{1}{2} (\nabla \phi)^2 =
-\frac{3}{2a}(\phi+\theta),
\end{equation}
where $\theta=2\Phi/3a^3H^2$ and $\Phi$ is the gravitational
potential. After some more algebra the system again becomes a
Schr\"{o}dinger-like wave equation, but in $a$ rather than in $t$
and using $\psi^2=\chi$, such that
\begin{equation}
i\nu \frac{\partial \psi}{\partial a} = -\frac{\nu^2}{2} \nabla^2
\psi + V\psi + P\psi~, \label{eq:scrap2}
\end{equation}
with $V=\phi+\theta$ and $P$ as before.

\section{The Origin of Spatial Intermittency}

Many types of non-linear system display a time-evolution
characterized by the word ``intermittency''. While linear Gaussian
processes involve fluctuations that are symmetric about their mean
value, non-linear processes typically have highly skewed
distributions. In the context of time series, intermittent
processes often have long quiescent periods punctuated by bursts
of intense activity. In the spatial domain, intermittent processes
are ones in which isolated regions of high density are separated
by large voids; see Shandarin \& Zel'dovich (1989).

One particular aspect of galaxy clustering that has received some
attention over the years has been the property that the one-point
probability distribution of density fluctuations $p(\rho)$ appears
to have a roughly lognormal form, i.e. $\log \rho$ has a roughly
normal distribution (Coles \& Jones 1991). The lognormal is a
prime distribution producing intermittency, and was discussed in a
pioneering paper by Kolmogorov (1962).  Although in a qualitative
sense the application of the concept of intermittency to
large-scale structure seems plausible, a quantitative description
of how it arises is not easy to obtain. Drawing on ideas discussed
by Zel'dovich et al. (1985, 1987), Jones (1999) suggested an
analytical model for the cosmological context.On a simple level,
this is quite easy to understand. If one has a linear process such
that the output $Y$ is constructed by co-adding a large number of
independent contributing processes $X_i$,
\begin{equation}
Y=\sum_{1=1}^{N} X_i
\end{equation}
as $N\rightarrow \infty$ then the central limit theorem guarantees
that $Y$ is Gaussian as long as the $X_i$ have finite variance. If
one takes instead a multiplicative process of the form
\begin{equation}
Y=\prod_{i=1}^{N} X_i
\end{equation}
with the $X_i$ still independent then the same theorem suggests
that $\log Y$ should be normal as $N\rightarrow\infty$. Lognormal
distributions consequently arise naturally in random
multiplicative processes, such as those involving fragmentation or
coagulation.

Whatever the details of its origin, it is now established that
this property has an interesting connection with the scaling
properties of moments of the probability of the distribution.
Taking a generic random variable $X$, such that the distribution
of $X$ within cells of side $L$ is denoted $p(X;L)$, then the
$q$-th moment at a given value of $L$ is said to display scaling
if
\begin{equation}
\langle X^q \rangle_L = \sum p(X;L)X^q \propto L^{\mu(q)}.
\end{equation}
This means that different powers $q$ of the density field vary as
a different power of the coarse-graining scale $L$. The function
$\mu(q)$ is called the {\em intermittency exponent}, and it can be
extracted from observations. Jones, Coles \& Martinez (1992)
showed that observations suggest a roughly quadratic dependence of
$\mu(q)$ upon $q$ and that this is related to the underlying
near-lognormal form of the density fluctuations. A set displaying
the form (41) is usually termed a multifractal; see Paladin \&
Vulpiani (1987) for general discussion.

We know that the distribution of density fluctuations is not
exactly lognormal. The intermittency exponent can be written in
the form
\begin{equation}
\mu(q)=-(q-1)D_q,
\end{equation}
where the $D_q$ are scaling dimensions ($D_2$ for example is the
correlation dimension). We know that $D_q\propto q$ for a
multifractal model whereas perturbative methods suggest a simpler
form of scaling such that $D_q=D_0$ is constant, typical of a
monofractal.

One aspect of this is that the hierarchy of correlation functions
that describe a lognormal distribution display Kirkwood (1935)
scaling, while it appears from numerical $N$-body simulations that
cosmological fluctuations display a different hierarchical form.
For a discussion of the relationship between lognormal  and
hierarchical scaling, see Coles \& Frenk (1991).

It is within the overall framework of the fluid model that Jones
(1999) sought to understand the observed intermittency of the
large-scale structure of the Universe. Using the velocity
potential introduced above, he first introduces an effective
Bernoulli equation for the flow:
\begin{equation}
{\partial \phi_v \over \partial t} - {(\nabla \phi_v)^2\over
2a^2}=\phi,
\end{equation}
where $\phi$ is the actual gravitational potential. This equation
neglects terms involving pressure gradients as mentioned above. To
cope with shell-crossing events, Jones (1999) introduces an
artificial viscosity $\nu$ by adding a term to the right-hand-side
of this equation:
\begin{equation}
{\partial \phi_v \over \partial t} - {(\nabla \phi_v)^2\over
2a^2}=\phi+{\nu\over a^2} \nabla^2\phi_v.
\end{equation}
The viscosity is introduced to prevent the particle flow from
entering the multi-stream region by causing particles to stick
together at shell-crossing. This is also used in an approach
called the adhesion approximation (Gurbatov, Saichev \& Shandarin
1989). Using the Hopf-Cole transformation $\phi_v=-2\nu\log
\varphi$ and defining a scaled gravitational potential via
$\phi=2\nu\epsilon$ we can write the Bernoulli equation as
\begin{equation}
{\partial \varphi \over \partial t} = \nu \nabla^2 \varphi +
\epsilon({\bf x}) \varphi. \label{eq:heat}
\end{equation}
This is called the random heat equation, because of the existence
of the spatially-fluctuating potential term $\epsilon({\bf x})$.
The gravitational potential changes very slowly even during
nonlinear evolution (Brainerd, Scherrer \& Villumsen 1993; Bagla
\& Padmanabhan 1994), so Jones (1999) assumes that it can be taken
as constant and to be Gaussian distributed. An approximate scaling
solution to (\ref{eq:heat}) can then be found using a path
integral adapted from that normally used in quantum physics
(Feynman \& Hibbs 1965); see below for more details. In this
approximation, the function $\varphi$ has a lognormal distribution
(Coles \& Jones 1991). We refer the reader to Jones (1999) for
details; see also Zel'dovich et al. (1985, 1987).

This model is one of the few attempts that have been made to
understand the non-linear behaviour of the matter distribution
using analytic methods. Although not rigorous it surely captures
the essential factors involved. It does, however, suffer from a
number of shortcomings. First, the approach does not follow
material beyond the shell-crossing stage. Second, the viscosity
$\nu$ that is needed does not have properties that are very
realistic physically:  it can depend neither on the density $\rho$
nor the position ${\bf x}$. Moreover, in the final analysis Jones
(1999) takes the limit $\nu\rightarrow 0$, so it cancels out
anyway. One is tempted to speculate that its introduction may be
unnecessary. Third, the function $\varphi({\bf x},t)$ that emerges
from equation (\ref{eq:heat}) is not the desired density
$\rho({\bf x}, t)$ nor does it bear a straightforward relation to
the density. Finally, it is not clear how the motion of a
collisional baryonic component can be modelled within this
framework.

This formulation provides a useful complementary approach to many
techniques, including $N$-body methods. It also provides a new
light with which to study the Jones (1999) model. Widrow \& Kaiser
(1993) show using theoretical arguments and numerical simulations
that this system allows accurate numerical evolution of the system
beyond shell-crossing, so it does not have the ad hoc construction
needed by the Jones (1999) model to remedy this.

Second, no artificial viscosity is required. Equations
(\ref{eq:heat}) and (\ref{eq:schrod}) are of the same form, apart
from minor subtleties like the use of complex time coordinates.
The potential term is easily understood in (\ref{eq:schrod}), and
the wavefunction $\psi$ now has a straightforward relationship to
$\rho$. The upshot of this is that if one adopts the approximation
of constant gravitational potential one can use the same path
integral approach as described by Jones (1999). In a nutshell,
given some initial wavefunction $\psi({\bf x'}, t')$ one can
determine the wavefunction at a subsequent time $\psi({\bf x}, t)$
using
\begin{equation}
\psi({\bf x}, t)=\int K({\bf x}, t; {\bf x'}, t') \psi ({\bf x'},
t') d^3x',
\end{equation}
where the function $K({\bf x}, t; {\bf x'}, t')$ involves a sum
over all paths $\Gamma$ connecting the initial and final states
with $t>t'$:
\begin{equation}
K({\bf x}, t; {\bf x'}, t')=\int {\cal D}\Gamma \exp
[iS(\Gamma)/\hbar]\label{eq:Green},
\end{equation}
where ${\cal D}$ is an appropriate measure on the set of classical
space-time trajectories. For a particle moving in a potential
$V({\bf x}, t)=m\phi({\bf x}, t)$ the action $S$ for a given path
$\Gamma$ is given by
\begin{equation}
S(\Gamma)=\int_{\Gamma} \left[ {1\over 2} m \dot{x}^2 - m\phi({\bf
x}) \right] dt.
\end{equation}
Note the presence of the Gaussian field in equation (48) and hence
in the exponential of the integrand  on the right-hand-side of
equation (\ref{eq:Green}). To get an approximate solution to this
system we can follow the same reasoning as Zeldovich et al. (1985,
1987) and Jones (1999), ignoring time-varying terms, using the
Gaussian properties and counting the dominant contributions to the
path integral to deduce that the integral produces a solution of
lognormal form. This part of the argument is identical to that
advanced by Jones, except that the solution is for $\psi$ rather
than $\varphi$ and since $\rho$ is $|\psi^2|$ then one directly
obtains a lognormal distribution for the desired density
$\rho({\bf x}, t)$.

It should be stressed that, although the present approach clearly
provides a more elegant formulation of the problem, the deduction
of lognormality remains approximate; the lognormal is not the
exact solution to the system to either Jones' equation (45) or to
equation (\ref{eq:schrod}). How accurately this approximate form
applies is open to doubt and will have to be checked by full
numerical solutions. Interestingly, however, it is known to apply
quite accurately in quantum systems such as disordered mesoscopic
electron configurations (Janssen 1998). As mentioned above, the
Schrodinger approach yields a wavefunction $\psi$ which is
directly related to the particle density $\rho$ via
$\rho=|\psi|^2$. Such quantum systems also display lognormal
scaling for properties such as the conductance, which depends on
$|\psi|^4$. It is a property of the lognormal distribution that if
a random variable $X$ is lognormal, then so is $X^n$. In such
systems the role of the gravitational potential $\phi$ is played
by a potential that describes the disorder of a solid, perhaps
caused by the presence of defects. Such systems display {\em
localisation} at low temperature which is similar in some ways to
the original idea of Anderson location (Anderson 1958). The
formation of strongly non-linear structures by gravity is thus
directly analogous to the generation of localised wavefunctions in
condensed matter systems.

Finally, and perhaps most promisingly for future work, the
equation (\ref{eq:schrod}) offers a relatively straightforward way
of modelling the behaviour of collisional material. The addition
to the potential of a term of the form $\kappa |\psi|^2$ (with
$\kappa$ an appropriately-chosen constant), converts the original
equation (\ref{eq:schrod}) into a nonlinear Schr\"{o}dinger
equation:
\begin{equation}
i\hbar {\partial \psi\over \partial t}= - {\hbar^2 \over 2m}
\nabla^2\psi + m \phi({\bf x})\psi + \kappa |\psi|^2
\psi\label{eq:schrodnl}
\end{equation}
(Sulem \& Sulem 1999). This equation is now equivalent to those
that describe the flow of a barotropic fluid; see Spiegel (1980).
This system can therefore be used to model pressure effects, which
are otherwise only handled effectively using numerical methods
such as smoothed-particle hydrodynamical approximations (e.g.
Monaghan 1992). In the context of quantum systems, the nonlinear
term is used to describe the formation of Bose-Einstein
condensates (e.g. Choi \& Niu 1999).

\section{Discussion}
In this lecture I have sketched out an approach to the study
evolving cosmological density fluctuations that relies on a
transformation of the Vlasov-Poisson system into a
Schr\"{o}dinger-Poisson system. The transformation is not a new
idea, but despite the efforts of Widrow \& Kaiser (1993) it does
not seem to be well known in the astronomical community. The
immediate advantage of this new formalism is that it yields a
rather more convincing approach to understanding the origin of
spatial intermittency and approximate lognormality in the galaxy
distribution than that offered by Jones (1999). It also makes a
connection in the underlying physics with other systems that
display similar phenomena.

On the other hand, one must be aware of the approximations also
inherent in the present approach. The Schr\"{o}dinger equation is
not exact, and its usefulness as an approximate tool is restricted
by a number of conditions outlined by Widrow \& Kaiser (1993); see
also Spiegel (1980). Furthermore, the lognormal solution of the
system is a further approximation and may not be valid especially
in the strongly-fluctuating limit. Although it neatly bypasses
some of the problems inherent in the Jones (1999) model, the
nonlinear wave equation is by no means easy to solve in general
situations. Numerical methods will still have to be employed to
understanding other aspects of the evolution of cosmic structure
within this framework as indeed they are in other branches of
physics.

One particular issue worth exploring using this approach is to
understand the limits of the approach in strongly non-linear
situations. As it stands, the justification for the lognormal
approximation arises from the weakly non-linear behaviour of
collisionless matter moving in an almost constant potential field.
Taking into account the expansion of the Universe, the changing
gravitational potential, and the possible effects of matter
pressure within in the action formalism  may well reveal that a
different form of hierarchical scaling pertains in the strongly
non-linear regime.

It is perhaps worth mentioning some specific ideas of things that
could be done using this approach and for which there seem to be
clear benefits.
\begin{itemize}
\item {\bf Perturbation Theory.} Standard perturbation methods do
not guarantee a density field that is everywhere positive.
Re-casting cosmological perturbation theory in terms of $\psi$,
constructed so that $\psi^2=\rho$ can remedy this.
\item {\bf Gas Pressure.} Analytic techniques for modelling the
effects of gas pressure are scarce, even in relatively simple
systems such as Lyman-$\alpha$ absorption cloud (Matarrese \&
Mohayee 2002). The quantum pressure term (36) or alternative terms
such as in (56) allow flexibility to model gas behaviour at least
at a phenomenological level.
\item {\bf Shell-crossing.} Methods such as the Zel'dovich
approximation break down at shell-crossing, as described in
Section 2.6. Although the simple {\em ansatz} I have used in this
lecture does assume a unique velocity at every fluid location, it
is possible to construct more complex representations that allow
for multi-streaming (Widrow \& Kaiser 1993). Note also that no
singularities occur in the wavefunction at any time.
\item {\bf Reconstruction.} It is interesting to speculate that it
might be possible to use the unitary structure of quantum
mechanics in order to turn back the clock on evolved structure in
order to reconstruct parts of the cosmic initial conditions.
\end{itemize}

In general many techniques exist for studying the wave mechanics
of disordered systems, such as the renormalization group and
path-integral methods, few of which are used by cosmologists
working in this area. It is to be hoped that the introduction of
some of these methods may allow better physical insights into the
behaviour of non-linear structure formation than can be found
using brute-force $N$-body techniques.

\section*{Acknowledgments}

I wish to thank Roya Mohayee for discussions on some of the
material presented in this contribution.

\end{document}